\input harvmac

\def\title#1#2#3{\nopagenumbers\abstractfont\hsize=\hstitle
\rightline{hep-th/9603101}
\rightline{ITF-#1}
\bigskip
\vskip 0.7in\centerline{\titlefont #2}\abstractfont\vskip .2in
\centerline{{\titlefont#3}}
\vskip 0.3in\pageno=0}
\def\Bernd{\centerline{{\bf B.J. ~Schroers}\footnote{$^{\dagger}$}{e-mail:
{\tt schroers@phys.uva.nl} }}
\bigskip
\centerline{
Institute for Theoretical Physics}
\centerline{University of Amsterdam} 
\centerline{
Valckenierstraat 65, 1018 XE Amsterdam}
\centerline{The Netherlands}
\bigskip\bigskip}
\def\abs#1{\centerline{{\bf Abstract}}\vskip 0.3in
\baselineskip 12pt
{#1}}
\def\ack{\bigbreak\bigskip\smallskip\centerline{{\bf Acknowledgements}}
\nobreak}

\def\d{\dagger}

\def\bphi{\vec{\phi}}
\def\bCP{{\hbox{\bf CP}}}
\def\bC{{\hbox{\bf C}}}
\def\bZ{{\hbox{\bf Z}}}
\def\bR{{\hbox{\bf R}}}
\def\bF{{\hbox{\bf F}}}
\def\bV{{\hbox{\bf V}}}
\def\bW{{\hbox{\bf W}}}
\def\bQ{{\hbox{\bf Q}}}

\def\pmatrix#1{\left(\matrix{#1}\right)}

\lref\Nick{N.S. Manton, ``A remark on the scattering of BPS monopoles'',
Phys. Lett. {\bf  110B} (1982) 54
}
\lref\Schroers{  B.J. Schroers, ``Bogomol'nyi solitons in a gauged
$O(3)$ sigma model'', Phys.Lett. {\bf   356B } (1995) 291
}
\lref\Witten{ E. Witten, ``Phases of N=2 theories in two dimensions'',
 Nucl.Phys. {\bf  B403} (1993) 159
}
\lref\MP{D.R. Morrison and M.R. Plessner, ``Summing the instantons:
quantum cohomology and mirror symmetry in toric varieties'', Nucl.Phys.
{\bf B440} (1995) 279
}
\lref\toric{
M. Audin, {\sl The topology of torus actions on symplectic manifolds}
(Birkh\"auser, Basel 1991)
}
\lref\Raja{ R. Rajaraman, {\sl Solitons and instantons} (North-Holland,
Amsterdam 1982)
}
\lref\statmech{
N.S. Manton, ``Statitstical mechanics of vortices'',
 Nucl.Phys. {\bf B400} (1993) 624
\semi
P.A. Shah and N.S. Manton, ``Thermodynamics of vortices in the
plane'', J.Math.Phys. {\bf 35} (1994) 1171
\semi 
P.A. Shah, ``Statistical mechanics
of vortices on Riemann surfaces'', DAMTP/HEP 95-45 
}
\lref\Vacha{ T. Vachaspati and A. Achucarro, ``Semilocal cosmic strings'',
Phys.Rev. {\bf  D44} (1991) 3067
}
\lref\Gibetal{ G.W. Gibbons, M.E. Ortiz, F. Ruiz Ruiz and T.M. Samols,
 ``Semilocal strings and monopoles'',
Nucl.Phys. {\bf B385} (1992) 127
}
\lref\Trevor{ T.M. Samols, ``Vortex scattering'', Commun.Math.Phys.  {\bf 145}
(1992) 149
}
\lref\LS{R.A. Leese and T.M. Samols, ``Interaction of semilocal vortices'',
Nucl.Phys. {\bf  B396} (1993) 639
}
\lref\Hindmarsh{
M. Hindmarsh, ``Existence and stability of semilocal strings'', 
Phys.Rev.Lett. {\bf 68} (1992) 1263;
``Semilocal topological defects'', Nucl.Phys. {\bf  B392 }
(1993) 461
}
\lref\WardLeese{
R.S. Ward, ``Slowly moving lumps in the $\bCP^1$ model in (2+1) 
dimensions'', Phys.Lett. {\bf 158B} (1985) 424
\semi 
 R.A. Leese, ``Low-energy
scattering of solitons in the $\bCP^1$ model'', Nucl.Phys.
{\bf  B344} (1990) 33
}
\lref\bogvor{A. Jaffe and C. Taubes, {\sl Vortices and monopoles}
 (Birkh\"auser,
Boston 1980) 
}
\lref\CSHiggs{
J. Hong, Y. Kim and P.Y. Pac, ``On the multivortex solutions of the
abelian Chern-Simons-Higgs theory'',   Phys.Rev.Lett. {\bf 64} (1990) 2230
\semi
R. Jackiw and E. Weinberg, ``Self-Dual Chern-Simons vortices'',
 Phys.Rev.Lett {\bf 64}  (1990) 2234
\semi
 R. Jackiw, K. Lee and E.J. Weinberg, ``Self-Dual Chern-Simons Solitons'',
 Phys.Rev. {\bf D42} (1990)
3488.
}
\lref\supercond{
E. Witten, ``Superconducting strings'', Nucl.Phys. {\bf B249} (1985) 557
}
\lref\FI{
P. Fayet and J. Iliopoulos, ``Spontaneously broken supergauge symmetries
and Goldstone spinors'', Phys.Lett. {\bf 51B} (1974) 461
}
\lref\VS{A. Vilenkin and E.P.S. Shellard, {\sl Cosmic strings and 
other topological defects}  (Cambridge University Press, Cambridge 1994)
}
\lref\Shah{
P.A.  Shah, ``Vortex scattering at near-critical coupling'', Nucl.Phys.
 {\bf B429}
(1994) 259;
P.A. Shah, ``Phase transitions in a vortex gas'', 
Nucl.Phys. {\bf B438} (1995)
589 
}
\lref\Yang{ Y. Yang, ``A necessary and sufficient condition for the 
existence of multisolitons in a self-dual gauged sigma model'',
preprint 1995, submitted to Commun.Math.Phys. 
}
\lref\Gosh{
P.K. Ghosh and S.K. Ghosh, ``Topological and non-topological solitons
in a  gauged $O(3)$ sigma model with Chern-Simons term'' Phys.Lett. {\bf 366B}
(1996) 199
}
\lref\Lee{
K. Kimm, K. Lee and T. Lee, ``Anyonic Bogomol'nyi soliton in a 
gauged $O(3)$ sigma model'',  CU-TP-712, hep-th/9510141
}
\lref\Tigran{
K. Arthur and D.H. Tchrakian, ``Topological and non-topological self-dual
Chern-Simons solitons in a gauged $O(3)$ sigma model'', preprint 
DIAS-STP-96-07
}
\lref\MaxSkyr{  J. Gladikowski, B.M.A.G. Piette and B.J. Schroers,
``Skyrme-Maxwell solitons in (2+1) dimensions'', Phys.Rev. {\bf D53}
(1996) 844
}
\lref\martin{ J.M. Speight, ``Low-energy dynamics of a $\bCP^1$ lump on a 
sphere'', J.Math.Phys. {\bf 36} (1995) 796
}
\lref\LLM{
 C. Lee, K. Lee and H. Min, ``Self-Dual Maxwell-Chern-Simons 
Solitons'', Phys.Lett. {\bf 252B} (1990) 79
}
\lref\Kim{
C. Kim, ``Self-dual vortices in the generalized abelian Higgs model with
independent Chern-Simons interaction'', Phys.Rev. {\bf D47} (1993) 673
}
\lref\semiCS{
A. Khare, ``Semilocal Chern-Simons vortices'', Phys.Rev. {\bf D46} (1992)
R2287
}
\lref\Jens{
J. Gladikowski, ``Topological Chern-Simons vortices  in the gauged $O(3)$
sigma model'', preprint DTP 95-75, hep-th/9603056, 
to appear in Zeitschrift f\"ur 
Physik C
}

\title{96-05}
{The spectrum of Bogomol'nyi solitons} 
{in gauged linear sigma models}

\Bernd

\abs
\noindent  
Gauged linear sigma models with $\bC^m$-valued scalar fields and
gauge group $U(1)^d$, $d\leq m$,  have soliton solutions of 
Bogomol'nyi type if a suitably chosen potential for the scalar fields
is also included in the Lagrangian. Here  such  models are studied 
on $(2+1)$-dimensional Minkowski space.  If the dynamics of the 
gauge fields is governed by a Maxwell term the appropriate potential
is a  sum of  generalised Higgs potentials known as  Fayet-Iliopoulos
D-terms. Many interesting topological solitons of Bogomol'nyi type 
arise in models of this kind, including various types of vortices 
(e.g. Nielsen-Olesen, semilocal and  superconducting vortices) as 
well as, in certain limits, textures (e.g. $\bCP^{m-1}$ textures and 
gauged $\bCP^{m-1}$ textures). This is explained  and general results  
about the spectrum of topological defects both for broken and 
partially broken gauge symmetry are proven. When the dynamics 
of the gauge fields is governed by a Chern-Simons term instead of a  
Maxwell term a different scalar potential is required for the theory 
to be  of Bogomol'nyi type. The general form of that potential is 
given and a particular example is discussed.

\Date{March 1996}

\baselineskip 16pt
\newskip\normalparskip
\normalparskip = 4pt plus 1.2pt minus 0.6pt
\parskip = \normalparskip 
\newsec{Introduction}

 In the study of topological solitons, field theories of Bogomol'nyi
type occupy a special place.
 Mathematically such field theories
are characterised by the fact that their  soliton solutions 
may be obtained by solving first order differential  equations
(called the Bogomol'nyi equations) instead
of the second order Euler-Lagrange equations. 
A  further characteristic  property 
 is the existence of  static multisoliton solutions
made up of arbitrarily  placed single solitons.
Physically one may think of this property as  reflecting
the absence of static forces between well-separated single
solitons. It follows that,  in theories
of Bogomol'nyi type, the space of static multisoliton
solutions  is larger and more interesting than in 
generic field theories with
topological solitons. Defining 
  the moduli spaces $M_{N}$    to be 
the spaces of static soliton  solutions of soliton number  $N$, 
 one finds  for example 
that in theories of Bogomol'nyi type  the dimension of
$M_N$  increases linearly with  $N$.
Starting with  the proposal by Manton \Nick\ in the context
of magnetic monopoles, moduli spaces have become a powerful
tool for investigating the physics of several interacting
solitons in theories of Bogomol'nyi type . This program
is particularly advanced in the case of  vortices in the abelian
 Higgs model. Here moduli space techniques have  been used 
successfully  to study the 
interacting dynamics of several vortices \Trevor\
and, more recently,  the statistical mechanics of vortices
\statmech.

The purpose of this paper is to explore a  
certain  class of (2+1)-dimensional
field theories with  topological solitons of Bogomol'nyi type
which provide  a natural generalisation  of the 
abelian Higgs model.
In two spatial  dimensions, it is useful to distinguish two types of 
topological solitons: vortices and textures (depending on the 
context the latter are also often called lumps or baby-Skyrmions). 
Generally we may 
distinguish the two by the origin of their topological stability.
If the stability is due to a non-trivial  first
 homotopy group of the  vacuum manifold we speak of vortices,
if it is due  to a non-trivial  second homotopy group of the vacuum
manifold
we speak of textures (though care has to be taken in interpreting
these statements,  as  first pointed out in ref.  \Vacha\
and as we shall also see later). More specifically, vortices
  arise
in field  theories with  charged scalar fields
 when  a $U(1)$  gauge symmetry is spontaneously broken,
 whereas textures typically 
arise in non-linear sigma models.
Typical and much studied examples of such textures are the lumps
in $\bCP^m$ sigma models.

Remarkably, both  vortices and textures can  be studied 
in a unified manner in the  framework of   gauged 
linear sigma models (GLSM's)  with a judiciously chosen 
Higgs potential.  The  special potentials which are required  here 
 are known to supersymmetry theorists as Fayet-Iliopoulos D-terms,
and GLSM's with these potentials  have been studied
intensively in recent
years in the context of topological sigma models and 
string theory, see  for example refs.  \Witten \MP.
  It has also been
emphasised in that  context that, in suitable limits,
 GLSM's with  Fayet-Iliopoulos   D-terms  become non-linear sigma models
whose target spaces are  certain  special manifolds called  
toric varieties (of which   the  projective spaces $\bCP^m$ 
are particular  examples).  Thus there exists a 
unifying framework for studying textures and vortices of Bogomol'nyi
type, with a host of beautiful  mathematical results, whose
implications for the study of topological defects in 
(2+1) dimensional field theories
do not  appear to have been fully exploited.

However,  while there is significant overlap between
the questions studied by string theorists and   those  which  one might
ask in the context of (2+1)-dimensional field theories, there are 
also important differences. While in the context of topological
sigma models and string theory GLSM's have typically 
been studied on   compact Riemann surfaces, 
the field theorist would naturally study the models  on flat 
Minkowski space. Thus, in the (2+1)-dimensional case
 static fields are defined on (non-compact) $\bR^2$,
and it is natural also to  consider  time-dependent fields. This leads
to additional questions. On a  non-compact domain 
the convergence of integrals needs to be checked carefully.
This results in extra constraints which affect the spectrum
of Bogomol'nyi solitons. The inclusion of the time  coordinate
allows one to study  time-dependent fields and to 
consider gauge fields whose dynamics 
is governed by a Chern-Simons term. The latter is physically 
interesting because topological solitons in such theories
display anyonic statistics.

This paper begins with two general sections in which 
GLSM's  are introduced and general results about the 
spectrum of Bogomol'nyi solitons are derived. 
Most of the algebraic manipulations in these sections
are standard, but the parts of the analysis which 
deal with the non-compactness of $\bR^2$ appear to
be new.  In sect.  4  we show  how the   non-linear
 $\bCP^1$ sigma model   can be understood in a very precise way
 as a limiting case of the  semilocal
vortex model  in the 
context of GLSM's, and we explain how this limiting procedure  can be
generalised. Our contribution in  this section is mainly 
an expository one, linking work done in the study of topological
defects  with the mathematical framework of toric varieties.
In sect. 5   we study in detail  
a family of models with  two complex scalar fields
and gauge group $U(1)\times U(1)$.   
We show that the topological solitons 
in  this  family include, for various parameter
values and in various limits, superconducting vortices and 
gauged $\bCP^1$ lumps.  The latter are solitons in the 
gauged $O(3)$ sigma model introduced in \Schroers.  
The present paper  developed  out of an attempt better  to understand 
the mathematical  structure and physical interpretation of 
that model, and we will see that the  framework of  GLSM's with
Fayet-Iliopoulos D-terms provides a satisfactory understanding of both.
In sect.  6
 we  write down a general Lagrangian for GLSM's of Bogomol'nyi
type with Chern-Simons  terms for the gauge fields. Again we illustrate
the general results  in a particular model with two complex scalar fields.
Finally, the purpose of sect. 7  is to draw together the
rather diverse viewpoints which enter this paper and to 
 highlight some open questions.

\newsec{Gauged linear sigma models}

Perhaps the most natural way to introduce  GLSM's with Fayet-Iliopoulos
D terms
without invoking supersymmetry is to recall the manipulations
that lead to the establishment of the Bogomol'nyi equations for the 
abelian Higgs model. The basic fields in the  abelian Higgs model 
are a complex scalar and a $U(1)$ gauge field,
but as we shall see presently
Bogomol'nyi equations can still be established, {\sl mutatis
mutandis},  if we consider instead  a $\bC^m$-valued scalar field $w$
and gauge any subgroup of the maximal torus of the unitary group
$U(m)$ acting on $w$ in the fundamental representation.
The maximal torus of $U(m)$ is $m$-dimensional  and 
a choice of  generators  $t_a$, $a=1, ... m$, defines an 
isomorphism between
it and   $U(1)^m$.  Concretely the generators are  diagonal $m\times m$
matrices with integer entries, so we can write in components  
\eqn\qa{\eqalign{ 
t^a_{\alpha \beta}= Q^a_{\alpha}
\delta_{\alpha\beta}
\qquad (\alpha,\beta = 1,...,m, \hbox{ no sum over}\,\alpha). 
}}   
Introducing a   $U(1)$ gauge fields $A^a_{\mu}$
 for each of the generators we define the covariant derivative
\eqn\aaa{\eqalign{
D_{\mu}w = \partial_{\mu}w + i\sum_{a=1}^m t_a  A^a_{\mu} \, w, 
}}
and the curvatures
\eqn\aaa{\eqalign{ 
F^a_{\mu \nu} = \partial_{\mu} A^a_{\nu} -\partial_{\nu} A^a_{\mu}.
}}
Here  the space-time indices $\mu,\nu,.. $ run  over $\{0,1,2\}$.
In the following we shall exclusively work on 
(2+1)-dimensional Minkowski space with signature $(-,+,+)$, whose
points we denote by $x^{\mu}$. Occasionally we also use latin
indices $i,j,k ...\in \{1,2\}$ to label  the spatial components of 
$x^{\mu}$, or polar coordinates $(\rho,\theta)$  for  $(x_1,x_2)$.
Finally we should define a suitable ``Higgs potential''  for each
generator $t_a$ of the gauge group  $U(1)^m$. The appropriate potential
turns out to be
 the Fayet-Iliopoulos D-term $(R_a - w^{\d}t_aw)^2$ \FI, where $R_a$
is a parameter of dimension mass. There are a number of 
ways to think about this term, but we only point out as an aside
that in terms of the symplectic geometry of $\bC^m$, with symplectic
form $dw^{\d}\wedge d w$,  $H_a = R_a -w^{\d}t_aw$ is a 
 Hamiltonian for the
$U(1)$ action on $\bC^m$ generated by $t_a$. The interested reader
is referred to the book \toric\  for more details on this point
of view.

Thus  we can write down the general  Lagrangian density 
 which is the main subject of this paper.  In natural units
$c=\hbar=1$ it reads
 \eqn\Lag{\eqalign{
{\cal L} = -{1\over 2}
 (D_{\mu}w)^{\d}D^{\mu}w -
\sum_{a=1}^m {1\over 4  e_a^2} (F^a_{\mu\nu})^2 -
\sum_{a=1}^m {e_a^2\over 8} (R_a - w^{\d}t_a w)^2.
}} 
It depends on $m$ coupling constants $e_a$ of  dimension 
 (mass)$^{1\over 2}$,
and on the  $m$ parameters $R_a$.
  The dependence of the theory on these 
parameters is one of our main interests. Note in particular
that we can trivially  eliminate any  of the  gauge fields
$A^a_{\mu}$  by setting  them and the corresponding 
coupling constants $e_a$
 to zero.

We  are  mostly interested in static fields,  for which 
the energy functional has the  form
\eqn\aaa{\eqalign{ 
E ={1\over  2} \int d^2 x\, (D_1 w)^{\d} D_1w +  (D_2 w)^{\d} D_2w
  + \sum_{a=1}^m
{1\over e_a^2} (F^a_{12})^2
 +  \sum_{a=1}^m {e_a^2\over 4} (R_a -w^{\d}t_a w)^2.
}}
To ensure that the energy of a configuration is finite we also
impose the boundary conditions
\eqn\bounda{
\lim_{\rho \rightarrow \infty} (R_a - w^{\d} t_a w) =0 
}
\eqn\boundb{
\lim_{\rho \rightarrow \infty} D_iw =0.
}
Below we will state more precisely  how quickly these limits
should be attained. For now we proceed,  assuming that the 
decay of the fields at infinity is fast enough to justify
the following manipulations. Thus using 
 the  algebraic identity
\eqn\aaa{\eqalign{
E = {1\over 2}\int d^2 x |(D_1 \pm i D_2)w|^2 +
 \sum_{a=1}^m|{1\over e_a} F^a_{12} \pm {e_a\over 2}
 (R_a - w^{\d}t_a w)|^2
\cr
\mp {1\over 2} 
\int d^2x \, i (D_1w)^{\d}D_2 w -i (D_2w)^{\d}D_1w + \sum_{a=1}^m
 F^a_{12}
 (R_a - w^{\d} t_a w)
}}
and   integrating  by 
parts we   deduce  
\eqn\aaa{\eqalign{
E = {1\over 2} \int d^2 x \, |(D_1 \pm i D_2)w|^2 +
 \sum_{a=1}^m|{1\over e_a} F^a_{12} \pm& {e_a\over 2} 
(R_a - w^{\d}t_a w)|^2 
\cr  \mp &{1\over 2 } \sum_{a=1}^m R^a \int d^2 x\, F^a_{12},
}}
where we have also used  that 
\eqn\curvature{\eqalign{
w^{\d}(D_1D_2 -D_2D_1)w  = i\sum_{a=1}^m w^{\d} t_a w \, F_{12}^a.
}}
Finally  defining suitably normalised  magnetic fluxes
\eqn\chern{\eqalign{
\Phi_a = -{1\over 2\pi}\int d^2x \, F^a_{12}
}} 
and
\eqn\aaa{\eqalign{
T =  \sum_{a=1}^m R_a \Phi_a
}}
we deduce the inequality
\eqn\aaa{\eqalign{
E \geq \pi |T|.
}}
More precisely we have
\eqn\holbound{\eqalign{
E = \pi  T
}}
if and only if   the Bogomol'nyi equations hold
\eqna\hol
$$\eqalignno{
(D_1 +i D_2)w &=0 &\hol a  \cr
F_{12}^a  + {e^2_a\over 2} (R_a -w^{\d}t_a w)& = 0, &\hol b \cr
}
$$
and 
\eqn\antiholbound{\eqalign{
 E=- \pi T
}}
if and only if  the (anti)-Bogomol'nyi equations hold
\eqna\antihol
$$\eqalignno{
(D_1 -i D_2)w & =0&\antihol a \cr
F_{12}^a - {e^2_a\over 2}  (R_a -w^{\d}t_a w)  & =0.&\antihol b\cr
}
$$

The quantity $T$  deserves two further comments. 
First  note that, by  converting the surface integral
 into a line integral
round the circle $C$ at  spatial infinity  and 
 using the boundary conditions \bounda\ and  \boundb, 
$T$ can be written as
\eqn\winding{\eqalign{ 
T = {1\over 2 \pi i} \oint_C w^{\d} dw.
}}
The second comment is a {\sl caveat}. Since we are working on non-compact
$\bR^2$ the fluxes $\Phi_a$ are not necessarily integers.
Thus neither the  fluxes  nor $T$  have, in general, 
a clear topological meaning. Nonetheless  
the  magnetic fluxes are   
interesting quantities to consider because  they are   conserved
if one rules out infinite energy configurations. This follows 
from Faraday's
law of induction
\eqn\fara{\eqalign{
{d \Phi_a \over d t} =-{1\over 2 \pi}  \oint_C E^a_i dx_i,
}}
where $E^a_i = F^a_{0i}$ is the electric field of the $a$-th gauge field.
 The integral on the right hand side is only non-zero
if the electric field falls off  for large  $\rho$    no  faster than
$1/\rho$, which is precisely the condition for the  electric field to 
have infinite  energy.

\newsec{The Bogomol'nyi equations on $\bR^2$}

For   the rest of this paper we 
focus on the Bogomol'nyi equations \hol{a}\ and \hol{b}.
We are interested in finite energy solutions of these
equations, and we need to to  specify more 
carefully  the  boundary condition which ensure
that the  energy and flux integrals  written down
 in the previous section converge. 
For this purpose  it is convenient to introduce the complex notation
 $z= x_1+ix_2$,   $\partial_z = {1\over 2}(\partial_1 -i\partial_2)$
and $A_a = {1\over 2} (A^a_1 -iA^a_2)$.
Further  we define gauge potentials and curvatures
\eqn\iiii{\eqalign{
a_{\alpha} = \sum_{a=1}^m Q_{a \alpha} A_a \cr
f^{\alpha}_{12} = \sum_{a=1}^mQ_{a \alpha} F^a_{12}
}}
and  we introduce the 
  set of  parameters $r_{\alpha}$ through
\eqn\rR{\eqalign{
R_a  = \sum_{\alpha =1}^m Q_{a\alpha}  r_{\alpha} 
}}
(note that the matrix $\bQ$ with entries $Q_{a\alpha}$ is 
invertible by virtue of the  generators $t_a$ being independent).
The parameters $r_{\alpha}$ are convenient  for discussing  the 
spontaneous symmetry breaking in this model. The potential
\eqn\aaa{\eqalign{
 \sum_{a=1}^m {e_a^2\over 8}  (R_a -w^{\d}t_a w)^2
}}
vanishes if and only if 
  $w^{\d}t_a w = R_a$ for all $a$  or, equivalently if 
\eqn\aaa{\eqalign{
|w_{\alpha}|^2 = r_{\alpha}, \qquad \alpha = 1, ... ,m.
}}
Thus, in order to have  any chance of  finding finite 
energy solutions we must require 
\eqn\ral{\eqalign{
r_{\alpha} \geq 0, \qquad \alpha= 1,...,m.
}} 
Further we see that the $U(1) $ gauge group which 
rotates the phase of $w_{\beta}$  for some given 
$\beta \in \{1,...,m\}$ is spontaneously
broken if $r_{\beta} > 0$  but  unbroken if $r_{\beta}=0$.

In the new notation 
 the Bogomol'nyi equations \hol{a}\  and \hol{b}\
 become
\eqna\newhol
$$
\eqalignno{
(\partial_{\bar z} + i \bar{a}_{\alpha})w_{\alpha}&= 0&\newhol a \cr
f_{12}^{\alpha} +{1\over 2}  
\sum_{a,\beta =1}^m e^2_a Q_{a\alpha} Q_{a\beta}
(r_{\beta} - |w_{\beta}|^2) &= 0.&\newhol b\cr
}
$$
 These two first order equations imply one second order
equation for $w_{\alpha}$ as follows. At points where $w_{\alpha}\neq 0$
 the first 
of the above equations is  equivalent to
\eqn\aaa{\eqalign{
\bar a_{\alpha} = i \partial_{\bar z} \ln  w_{\alpha},
}}
which in turn implies 
\eqn\aaa{\eqalign{ 
f_{12}^{\alpha} = {1\over 2} \Delta \ln |w_{\alpha}|^2.
}}
To extend this equation to the whole plane one uses 
the $\bar \partial$-Poincar\'e lemma in a  standard fashion,
see e.g.  \bogvor. 
Supposing  that  $w_{\alpha}$  has zeros at
 $\{z_{\alpha, s}| s=1, ..., n_{\alpha}\}$ 
 and using  that 
 in two dimensions
\eqn\aaa{\eqalign{ 
\Delta \ln |z-z_s|^2 = 4\pi \delta(z-z_s)
}}
 one concludes 
\eqn\fluxx{\eqalign{ 
f_{12}^{\alpha} = {1\over 2} \Delta \ln |w_{\alpha}|^2
-2\pi \sum_{s=1}^{n_{\alpha}} \delta(z-z_{\alpha, s}).
}}
Combining this with \newhol{b}\
 we arrive at  the  promised second order 
 equation for  $w_{\alpha}$:
\eqn\elliptic{\eqalign{
\Delta \ln |w_{\alpha}|^2  + 
\sum_{a,\beta =1}^m e^2_a Q_{a\alpha} Q_{a\beta}
(r_{\beta} - |w_{\beta}|^2 ) = 4\pi
\sum_{s=1}^{n_{\alpha}} \delta(z-z_{\alpha, s}).
}}

This equation  should be thought of as a generalised vortex
equation, in the sense that in the case  $m=1$ and $t_1=1$
it reduces to the equation
\eqn\abhivort{\eqalign{
\Delta \ln |w|^2  + e^2(r-|w|^2)= 4\pi \sum_{s=1}^N \delta(z-z_s)
}}
(we have omitted the label $\alpha =1$ and written $N=n_1$),
which is the key to the study of Nielsen-Olesen
 vortices in  the abelian Higgs model at critical coupling. 
As mentioned in the introduction 
many   of the questions we  will address  in this paper
are motivated  by what is known about the abelian Higgs model;  it may
thus be useful briefly  to  summarise some salient results.

The first question that one needs to settle is that of existence 
and uniqueness of solutions. In the case of \abhivort\ this question
is completely answered in  \bogvor.
It   is shown there that no finite-energy solutions exist
if $r\leq 0$, but that for $r>0$,  \abhivort\  has a unique
finite energy solution for each   integer $N\geq 0 $ 
 and given zeros $z_s$, $s=1, ..., N$. 
It follows  that 
for given $N\geq 0$ (and  thus  fixed energy)  the Bogomol'nyi
equations in the abelian Higgs model 
have  an  $N$  complex parameter family of solutions, the $N$ 
complex  parameters
determining the positions of the zeros of the $w$-field up to permutations.
This  parameter space  is called the moduli space of $N$-vortices.
The moduli space is in fact  a differentiable manifold  
and has  a natural Riemannian metric, inherited
 from the field theory kinetic 
energy functional.
This  metric is crucial in the so-called moduli space to 
vortex  dynamics. In that scheme
the interacting  dynamics of $N$ vortices is approximated by geodesic
motion on the moduli space of $N$-vortices. 
  In  \Trevor\ it is shown that the 
Bogomol'nyi property of the field theory implies
that the  metric on the moduli space is K\"ahler. 
The K\"ahler property  together with symmetries is then used 
to compute the metric  explicitly in the case $N=2$;
by studying the geodesics of this metric much can be learnt
 about the interaction of two vortices.
 The K\"ahler property is
also crucial in the study of statistical mechanics  of vortices 
on Riemann surfaces in ref.  \statmech.

The example of  vortices in the abelian Higgs model shows 
paradigmatically how the Bogomol'nyi  property is 
the key to understanding difficult  and interesting 
aspects of  vortex physics, such as vortex interactions
and statistical mechanics of vortices. The family of theories
defined in this and the previous  section  also have  soliton solutions 
of Bogomol'nyi type  whose physical properties can  be 
studied with similar methods. In this paper we will not attempt 
to do this in full generality but instead go through some of 
the steps outlined above for a few  
specific model. However, to end this general section we show how
one can  deduce some general information about the spectrum of 
Bogomol'nyi solitons  directly from   \elliptic.

In particular we can now state the conditions for
a solution of the Bogomol'nyi equations to have finite energy.
The energy of solutions of the  equations \newhol{a}\ and \newhol{b}\
can be written  conveniently in terms of the fluxes 
\eqn\flux{\eqalign{
\varphi_{\alpha}= -{1\over 2\pi}\int d^2 x \, f_{12}^{\alpha}. 
}}
It is 
\eqn\newen{\eqalign{ 
E = \pi T = \pi \sum_{\alpha= 1}^m r_{\alpha}\varphi_{\alpha}.
}}
It thus  follows from \newhol{b}\ that
 all fluxes and the energy are well-defined  if
 $(r_{\alpha} -|w_{\alpha}|^2)$
is integrable over $\bR^2$ for all $\alpha$.
If $r_{\alpha} >0$   for {\it all}  $\alpha$ the  gauge
symmetry is completely broken. Then all scalar and gauge fields 
approach their vacuum values exponentially
\eqn\expo{\eqalign{
|w_{\alpha}|^2 \approx  r_{\alpha}+ C_{\alpha}
 e^{-m_{\alpha}\rho} \qquad \hbox{for large} \quad \rho,
}}
with $m_{\alpha}>0$ and $C_{\alpha}$ arbitrary constants, and there 
are no convergence problems. 
On the other hand  if $r_{\beta}=0$ for some (at least one)
 $\beta\in \{1,...,m\}$  then the corresponding 
scalar fields $w_{\beta}$ and (because of the coupling) possibly 
some other scalar  
fields   approach their vacuum values slower than exponentially.
The precise   asymptotic behaviour depends on the matrix
$\,(\sum_{a=1}^m e^2_aQ_{a\alpha}Q_{a\beta})$ which appears in \elliptic\
and which couples the various components of $w$;  we will
give a more detailed discussion of 
some special cases later in this paper.  Here we
note that
it is consistent with \elliptic\ for  $|w_{\alpha}|$
to approach the vacuum according  to a 
power law
\eqn\power{\eqalign{
|w_{\alpha}|^2 \approx r_{\alpha}+
C_{\alpha} \rho^{-2\tilde \eta_{\alpha}}\qquad \hbox{for large} \quad \rho.
}}  
In that  case  we impose 
\eqn\finflux{\eqalign{
\tilde \eta_{\alpha} > 1
}}
to ensure finiteness of the  fluxes and the  energy.   
Finally it is possible that  the power-law decay is  modified
by logarithmic terms \foot{This was pointed out to me by
Trevor Samols}. In particular  we should also allow for 
the asymptotic form
\eqn\loga{\eqalign{
|w_{\alpha}|^2 \approx  r_{\alpha}+
{ C_{\alpha} \over \rho^2 \ln^2 \rho }\qquad \hbox{for large} \quad \rho,
}} 
which  (for suitable $e_a$ and $Q_{a\alpha}$) is 
 consistent  with \elliptic\  and the finite energy requirement.

We can obtain explicit formulae for the fluxes from \fluxx\  as follows.
Using Stokes's theorem  we  first  obtain
\eqn\parts{\eqalign{
\int d^2x\, \Delta\ln|w_{\alpha}|^2 = 2\pi \lim_{\rho \rightarrow \infty}
\rho {\partial\over \partial  \rho} \ln  |w_{\alpha}|^2 . 
}} 
Then    using  $\ln(r_{\alpha} + \epsilon) \approx 
\ln r_{\alpha} + \epsilon/r_{\alpha}$ if $r_{\alpha}\neq 0$ and $\epsilon$
small,  and integrating \fluxx\ 
we  deduce   
\eqn\fluxcona{\eqalign{
 r_{\alpha} >0 \, \, \Rightarrow \,\,
\varphi_{\alpha} = n_{\alpha}\in \bZ^{\geq 0} 
}}
regardless of how $|w_{\alpha}|^2$ approaches $r_{\alpha}$.
If on the other hand $r_{\alpha}= 0$  we know that $|w_{\alpha}|^2$
tends to zero according to \power\ or \loga. Then the result of the
integration can be summarised in 
\eqn\fluxconb{\eqalign{
r_{\alpha} =0 \,\, \Rightarrow \,\,
\varphi_{\alpha} = n_{\alpha} + \eta_{\alpha}, 
\quad n_{\alpha} \in \bZ^{\geq 0} \quad  \hbox{and}\quad  
 \eta_{\alpha} \in \bR^{\geq 1}.
}}
The real number $\eta_{\alpha}$
equals $\tilde \eta_{\alpha}$ if 
   $|w_{\alpha}|^2$    decays according to the power law \power\  
 and is 1 if $|w_{\alpha}|^2$ approaches
zero according to \loga.

The above formulae show that  the flux $\varphi_{\alpha}$ 
counts the zeros  of $w_{\alpha}$ with multiplicity. If 
$r_{\alpha} > 0$,  all zeros are  at finite $\rho$ and have
integer multiplicity. If $r_{\alpha} =0$ then $w_{\alpha}$
has a zero at infinity  whose multiplicity is  $\geq 1$
but not necessarily integer.
(One shows similarly that for  
 square-integrable  solution of eqs. \antihol{a}\
and \antihol{b},  $\varphi_{\alpha} \leq 0$ if $ r_{\alpha} >0$ and 
$\varphi_{\alpha} \leq- 1$  if $r_{\alpha} =0$.)
A  further useful condition can be deduced  from  \newhol{b}\
if $r_{\beta}= 0$ for some $\beta \in\{1,...,m\}$.
It  then follows that 
\eqn\aaa{\eqalign{
\sum_{a=1}^m (\bQ^{-1})_{\beta a }{F_{12}^a \over e_a^2} \geq 0
}}
so that 
\eqn\powercon{\eqalign{  
\sum_{a=1}^m (\bQ^{-1})_{\beta a }{\Phi_a \over e_a^2} \leq 0.
}}
Here the equality holds only if $w_{\beta}$ vanishes everywhere.

\newsec{Semilocal vortices,  $\bCP^1$ lumps and toric varieties }

For specific calculations we will mostly concentrate on the 
GLSM with two scalar fields $w_1$ and $w_2$ in this paper.
Our favourite choice of generators of $U(1)$ subgroups of 
the maximal torus in this case is 
 \eqn\generators{\eqalign{ 
t_1 = \pmatrix{1 & 0 \cr 0 &1} \qquad 
t_2 = \pmatrix{1 &0 \cr 0 & -1}
}}
(we will mostly  omit writing the identity matrix $t_1$ in
the following formulae). Then, to avoid  crowded notation, we 
 write $A_{\mu}$ and $B_{\mu}$ for the gauge fields
$A^1_{\mu}$ and $A^2_{\mu}$, $A$ and $B$ for the corresponding
complex fields $A^1$ and $A^2$,
 and $F_{\mu\nu}$ and $G_{\mu\nu}$
for the corresponding field strengths.

In this section we restrict attention to the case where
one of the gauge fields is set to zero. 
Specifically  consider the 
 case $e_2=0$ and $B_{\mu}=0$. 
The resulting model
has been much  studied in the recent literature and has  vortex 
solutions known  as semilocal vortices, see refs. 
 \Vacha, \Gibetal\ and \LS.
One interesting aspect of  this model   is that
it contains stable vortices although
the vacuum manifold, defined as the submanifold of $\bC^2$ 
where  the potential $(R_1- w^{\d}w)$  vanishes, is
a three-sphere (provided  $R_1>0$) 
and therefore has trivial first homotopy group.
The reason why  there are nonetheless 
stable vortex solution, first explained in \Vacha,
is that in addition to condition \bounda\
we have the condition
\boundb\ and this  forces  $w$  to lie both on the 
vacuum manifold and  on a gauge orbit
of the gauged $U(1)$ at spatial infinity.
 Thus, in terms of spatial polar coordinates
$(\rho,\theta)$   $w$ has to be  of the form
 $(w_1,w_2) = (c_1 e^{iN_1\theta}, c_2 e^{i N_1\theta})$ for  large $\rho$
with $N_1$ some integer and 
 $c_1$ and $c_2$ complex constants satisfying $c_1^2 + c_2^2 = R_1$.
The point is that the gauge group acts without fixed points
on the vacuum manifold, so that gauge orbits are necessarily
 loops and never just a point. The integer $N_1$ is the degree of  the map
 $w_{|\rho=\infty}$  from the circle at spatial infinity into
one of these gauge orbits.
Evaluating the formula  \winding\ for the topological
lower bound in this case we find $T=R_1 N_1$, showing that 
vortex solutions of the  Bogomol'nyi equations with $N_1\neq 0$
cannot decay into the vacuum.

It has  also been
 observed  in the literature  \Hindmarsh\
that there is a close connection between semilocal vortices 
and  topological solitons  in the $\bCP^1$  (or $O(3)$) sigma model.
This connection is usually discussed in  terms of energy scales,
with the sigma model being thought of as a low-energy
effective theory of the vortex model.  However, from  the 
present point of view it is more convenient  to keep the 
energy under consideration fixed and vary the parameter
$e_1$ (recall that this has dimensions (mass)$^{1\over 2}$). In fact
it is easy to see that in 
 limit $e_1 \rightarrow \infty$
the semilocal vortex model reduces to the $\bCP^1$ model.
To keep the energy 
finite we simultaneously  impose the  constraint
\eqn\consta{\eqalign{ 
R_1 - w^{\d} w = 0.
}}
Then, since  the kinetic term of the gauge field $A_{\mu}$  disappears
form the Lagrangian  
in the limit  $e_1 \rightarrow \infty$ we can eliminate the 
gauge field  altogether form the energy functional via the equation
\eqn\ao{\eqalign{
{\partial {\cal L}\over \partial A_{\mu}} = 
{1\over 2 i}( (D_{\mu} w)^{\d} w - w^{\d}D_{\mu}w) =0,
}}
which implies
\eqn\AAA{\eqalign{ 
 R_1  A_{\mu} = i w^{\d}\partial_{\mu} w.
}}
It is a standard result, see e.g. \Raja, 
 that the resulting model
is equivalent to the $\bCP^1$ sigma model. Geometrically,
the condition \consta\ forces $w$  to lie on a three-sphere
$S^3$, and the appearance 
 of the covariant derivative with 
the gauge potential given by \AAA\
means that the Lagrangian depends on $w$ only up to 
an overall phase. Thus,  defining the equivalence relation $\sim$
via
$(w_1,w_2)\sim e^{i \chi}(w_1,w_2)$, $\chi\in [0,2\pi)$,
  the limit $e_1 \rightarrow \infty$ leads 
to the  non-linear sigma model with target space
$(S^3/\sim)\,\,\cong  \bCP^1 \cong  S^2$. For later use, we note
that this can be made explicit by introducing the
$\bCP^1$-valued  field
\eqn\uuu{\eqalign{
u = {w_2\over w_1} 
}}
or  the $S^2$-valued field $\bphi= (\phi_1,\phi_2,\phi_3)$, defined
via
\eqn\pphi{\eqalign{
\phi_l = w^{\d}\tau_l w,  \qquad l=1,2,3,
}}
where $\tau_1,\tau_2$ and $\tau_3$ are the  three Pauli matrices.
The Lagrangian density  can then be written in terms of $u$ or $\bphi$,
see  ref. \Raja.

Note also that the topological character of the 
topological defect changes as we take the limit $e_1\rightarrow \infty$.
For the semilocal vortex 
we have already interpreted the topological bound $T$  as 
($R_1$ times)
the  degree of the $w$  viewed as a mapping from  the circle at 
 spatial infinity to a gauge orbit on the vacuum manifold.
 However, in the limit $e_1\rightarrow \infty$, points on gauge orbits
should be identified, so that now $w$ maps the entire circle
at spatial infinity into on point on the  target space $\bCP^1$.
This allows us to regard $u=w_2/w_1$  as a map from compactified
space $\bR^2 \cup \{\infty\}\cong  S^2$
 to $\bCP^1$.
The flux 
$\Phi_1=-1/2\pi \int d^2x F_{12}$, with $F_{12}$ computed from
\AAA,  equals  the degree of that map.  Thus, in the general
terminology of the introduction  topological solitons in 
the $\bCP^1$ model are textures. However, in this particular
context  they are more usually called lumps.

The moduli spaces of both  semilocal vortices and  $\bCP^1$  lumps   
have been studied in some detail.  In \Gibetal\ it is shown that 
the moduli space of semilocal vortices  with magnetic flux number
$N$ is diffeomorphic to the space of  polynomials  of the form  
$P_N(z) =
z^N + \sum_{s=0}^{N-1} a_s z^s$ and $Q_N(z)=\sum_{s=0}^{N-1} b_s z^s$.
The coordinatisation of that space in terms of the complex
coefficients
$a_s$ and $b_s$, $s=0,...,N-1$ shows that it  can be identified 
with $\bC^{2N}$. The translation from a point in the moduli space
to an actual field configuration, however, is best done in terms 
of the zeros of $P_N$ and $Q_N$. They are also the zeros  of the 
scalar fields $w_1$ and $w_2$.

In the $\bCP^1$ model, a lump of  degree $N$
 is a  rational  function on $\bR^2$  of degree $N$ which tend to 
zero at infinity. Explicitly such a function can again 
be written in terms
of the polynomials $P_N$ and $Q_N$
\eqn\rat{\eqalign{
u(z)  = {Q_N(z) \over P_N(z)},
}}
but now we have to require in addition that $P_N$ and $Q_N$
have no common zeros, or equivalently that 
the   resultant of $P_N$ and $Q_N$ 
is non-vanishing. Writing $R_N$ for the set where the resultant does
vanish  we conclude that the moduli space
of degree $N$ lumps  is  $\bC^{2N} - R_N$. 
The interpretation of the moduli is now quite different from 
the vortex case. Writing for example a single lump configuration
as $b/(z-a)$, the complex number $a$ is 
the lump's position in $\bR^2$  and  the modulus and phase   of  the 
complex number $b$ are its size and orientation respectively.
Note that $b$ has to be non-vanishing and that 
in the limit $|b| \rightarrow 0$ the lump becomes infinitely 
spiky. These features generalise to  
  lumps of degree $N$.  The moduli encode information
about  internal  as well as  position  degrees of freedom.
The condition of the non-vanishing resultant removes precisely
those points from the moduli space  which correspond to infinitely
spiky configurations.

As mentioned earlier, the Riemannian metric which the 
 moduli spaces  inherit  from the field theory kinetic energy
is crucial  
in the moduli space  approximation to soliton dynamics.
In ref.  \WardLeese\  this approximation is   applied to lumps
in the  $\bCP^1$
model and it turns out  that the 
 moduli  space  metric  has
two problematic features in this case (see also \martin\ for 
further discussion of this point). The metric  is
not finite   because changes in the 
coefficient $q_{N-1}$ require infinite kinetic energy. Furthermore
it is not 
complete: there are geodesics which reach infinitely spiky
configurations in finite time.  By contrast the metric
on the  moduli 
space for semilocal vortices , studied in ref.  \LS\  in the case $e_1 =1$, 
is finite and 
complete. Thus  the moduli space of semilocal vortices can be thought of 
in a very precise sense as a regularised  version of the 
moduli space of $\bCP^1$ lumps, to which it tends in the 
limit $e_1 \rightarrow \infty$.

Let us briefly consider the case where only 
 the $U(1)$-factor  generated by
$t_2$ is gauged. Thus we set $e_1=0$ and $A_{\mu}= 0$.
 The Bogomol'nyi
equations \hol{a} and \hol{b} are  this case:
\eqna\silly
$$
\eqalignno{
(\partial_{\bar z} + i \bar{B})w_1&= 0& \cr
(\partial_{\bar z} - i \bar{B})w_2&= 0&\silly a \cr
G_{12} + {1\over 2} ( R_2 -|w_1|^2 + |w_2|^2)  &= 0.&\silly b\cr
}
$$
The energy of  a solution of these equations is $E=R_2\Phi_2$.
Thus, if $R_2 < 0$ and $E\neq 0$ we necessarily have 
 $\Phi_2  <  0$. However, it then follows from 
a standard vanishing theorem (a line bundle of  negative
degree cannot have a non-zero holomorphic section) that $w_1 =0$.
Similarly if $R_2 > 0$ we deduce that for any solution 
with non-vanishing energy $w_2 =0 $
 In either case the vortex solutions of this 
model are just embedded Nielsen-Olesen vortices.  In the 
limit $e_2 \rightarrow \infty$ we again obtain a non-linear 
sigma model, but this time the target space is $\{(w_1,w_2)\in \bC^2
|R_2 = |w_1|^2 -|w_2|^2\}/\sim$, where $\sim$ is the equivalence 
relation $(w_1,w_2)\sim (e^{i\chi}w_1, e^{-i\chi}w_2)$. This space
is known in mathematics  as the  weighted projective space
 $\bCP^1_{(1,-1)}$ and is isomorphic  to $\bC$. Bogomol'nyi
solitons in the corresponding  non-linear sigma model  
are holomorphic maps from $S^2$ to  $\bCP^1_{(1,-1)}$.
However, such maps  are necessarily  constant maps. 
Thus there are no  non-trivial textures of Bogomol'nyi type in 
this model.

The  phenomena encountered in this section form part of 
a very general story, much discussed in mathematics and string
theory, see refs.  \Witten\  and \MP.
 Briefly, it goes as follows. Returning to the general
notation of sect. 2, consider taking the limit 
\eqn\limi{\eqalign{ 
 e_a \rightarrow \infty,  \qquad a \in I
}}
and simultaneously imposing
\eqn\constraint{\eqalign{
(R_a- w^{\d} t_a w)  = 0 , \qquad a \in I
}} 
for some 
subset of indices $I\subset \{1, ...,m\}$ which we can without
loss of generality take to be $I=\{1, ..., d\}$, $d < m$.
Assume that the $R_a$ are such that  the entire gauge symmetry
is spontaneously broken.
Also let us at first restrict our attention to 
the situation where the 
couplings and gauge fields labelled by the complementary
indices are set to zero
\eqn\comp{\eqalign{
e_a = 0, \qquad A_{\mu}^a =0 \qquad \hbox{for}\quad  a = d+1,...,m.
}} 
Defining the currents
\eqn\current{\eqalign{
j_{\mu}^a = {1\over 2i}\left( (D_{\mu}w)^{\d}t^a w -
w^{\d}t^a D_{\mu}w\right), 
}}
 the equations of motion  for the gauge fields $A_{\mu}^a$, $a\in I$
 are then
\eqn\elimaa{\eqalign{
j_{\mu}^a = 0 \quad \hbox{for} \quad a \in I
}}
which one can  solve explicitly  for $A_{\mu}^a$, $a\in I$:
\eqn\elim{\eqalign{
\sum_{b=1}^d w^{\d} t_a t_b w \, A^b_{\mu} = 
i w^{\d} t_a \partial_{\mu} w. 
}}
The gauge fields  $A_{\mu}^a$, $a\in I$,  can thus be eliminated,
but the Lagrangian \Lag\ is still invariant under
transformations $w(x)\rightarrow e^{i\chi^a(x) t_a}w(x),
 \quad a\in I,\quad \chi_a \in [0,2\pi)$.
Then defining the equivalence relation 
\eqn\torus{\eqalign{
w\sim e^{i \chi^a t_a }w,\qquad a\in I, \qquad \chi_a \in [0,2\pi)
}}
 the fields $w$  may be thought  of as 
taking values in the non-linear space  
\eqn\aaa{\eqalign{
Z = \{w\in \bC^m| w^{\d}t_a w = R_a, a \in I\}/\sim.
}}

The combined operation of  imposing \constraint\ and dividing
by the  action \torus\ of the torus  $U(1)^d$ is called the 
symplectic quotient of $\bC^m$ by $U(1)^d$ and is often written
$\bC^m // U(1)^d$. In fact what we are looking at here is 
a very special sort of symplectic quotient. With the original space
being $\bC^m$ and the group action being a $U(1)^d$-action the 
resulting quotient is  a  so-called toric variety of complex 
dimension $m-d$.
 The complex projective
space  and  the weighted complex projective space  which we 
encountered above are special examples of toric varieties. 
Toric varieties are  naturally  K\"ahler manifolds,
and their  topology (which depends on the values of the $R_a$)
 is well-studied. In particular it is known
that under some restriction on the $Q^a_{\alpha}$ \qa\
the quotient space $Z$ is compact. This  holds for example
if for some $a\in I$ all the $Q_{\alpha}^a$ are positive.
(For more general conditions see \toric.)
Moreover, the second homology group
$H^2(Z,\bR)$ is $d$-dimensional for generic values of $R_a$
 (note that as long as 
$d<m$, 
the real dimension of $ Z$ is  $ \geq 2$). 

In analogy to  our discussion of the $\bCP^1$ sigma model  
static fields which obey the condition \boundb\
can  here be regarded as  maps from $\bR^2 \cup \{\infty\} \cong S^2$ 
 into $Z$.  Such  maps  can be classified 
topologically by their
multi-degree, the integral of the pull-back of the $d$ generators
of $H^2(Z,\bR)$. This generalises the observation made 
 in the simple case of the $\bCP^1$ sigma model. For every
gauge field which we eliminate by taking the 
corresponding coupling constant  to infinity 
 the topological meaning
of  the magnetic 
flux number changes. If 
before elimination   this number 
  counts the number of times the  circle at spatial
infinity is wrapped round a certain loop in the vacuum manifold
then, 
after elimination,  it counts the number of times  compactified 
space get wrapped around a certain generator of the  second 
homology of the (now  non-linear) target space.

Note that the  argument in the last paragraph depends crucially
on the fact  that  all gauge fields are  either eliminated
through the limit \limi\ or set to zero \comp. Only in this 
case does the boundary condition \boundb\ allow us to  identify
points at infinity to one point.  However, it is also
interesting to consider the mixed situation, where some 
gauge fields are eliminated and others remain as dynamical
fields. This leads to gauged non-linear sigma models,
where the geometric 
interpretation of the magnetic fluxes $\varphi_{\alpha}$
is more subtle. We shall
 see this in a particular example in the next section.

\newsec{Topological solitons in $U(1) \times U(1)$ gauge theory}

We now have all the ingredients necessary for the study
of the  case where  the entire maximal torus of $U(2)$
is gauged. There are eight independent  parameters in this model:
the real numbers 
$e_1,e_2,R_1,R_2$ and the integer  entries  in the generators 
$t_a$ :
\eqn\generatorsa{\eqalign{ 
t_1 = \pmatrix{Q_{11} & 0 \cr 0 &Q_{12}} \qquad 
t_2 = \pmatrix{Q_{21} &0 \cr 0 & Q_{22}}.
}}
 It is instructive
to write out the Lagrangian density in components for this model
\eqn\suplag{\eqalign{
 {\cal L}& = -{1\over 2}| D_{\mu}w_1|^2 - {1\over 2}| D_{\mu} w_2|^2
- {1\over 4 e_1^2} F_{\mu \nu}^2 - {1\over 4e_2 ^2} G_{\mu \nu}^2  
\cr
&- {e_1^2\over 8}(R_1 -Q_{11}|w_1|^2 - Q_{12} |w_2|^2)^2 -
 {e_2^2\over 8}(R_2 - Q_{21}|w_1|^2 - Q_{22}|w_2|^2)^2.
}}
This Lagrangian density is of the type studied first by Witten 
in the seminal paper \supercond\ and later by many others
(see the book \VS\ and references therein) as a model
for  bosonic superconducting vortices. 
In that context  the discussion is usually conducted in terms 
of the gauge fields $a_{\mu} = Q_{11}A_{\mu} + Q_{21}B_{\mu}$ and 
$ b_{\mu} = Q_{12} A_{\mu} + Q_{22} B_{\mu}$
which couple only  to the scalar fields $w_1$ and $w_2$ respectively.
Writing $U(1)_a$  and  $U(1)_b$ for the gauge groups
 which rotate the phases  of $w_1$ and $w_2$  respectively
the  basic observation can be stated 
as follows. By adjusting parameters 
suitably one can arrange for  one of the two gauge groups, say $U(1)_a$, 
  to be  spontaneously
broken  while $U(1)_b$  remains unbroken. Thus $U(1)_b$ may
be interpreted as the gauge group of electromagnetism (there is 
a {\sl caveat} which we explain below).
Then there is a range of parameters
for which the  model has
 $U(1)_a$-vortex solutions in whose core the Higgs field $w_2$ of 
 $U(1)_b$ is non-zero, thereby breaking the electromagnetic gauge group
there. Witten showed that under these circumstances  the core of the 
vortex becomes superconducting.

The Lagrangian density \suplag\ is  more general than
  the Lagrangians usually studied in the context of superconducting 
vortices because it allows for interactions between the 
 gauge fields $a_{\mu}$ and $b_{\mu}$ (through the curvature
terms) but it also has  many   special properties
because  it of   Bogomol'nyi type. 
One general feature of Bogomol'nyi solitons in gauge theories
with scalar fields  is 
the cancellation between  the static forces mediated by the  gauge
fields and those mediated by  the scalar fields. 
Such a cancellation is of course only possible if  the 
scalar and gauge fields are either both massless or both
massive.  Thus we should expect  on general grounds that 
in our model, too, the  scalar field $w_2$ should acquire
a long range component when $U(1)_b$ is unbroken. In fact
it is easy to check this explicitly. 
By the  general condition stated after  eq. \ral,  $U(1)_b$
is unbroken if  and only if 
 $r_2=0$, which is equivalent to 
 $Q_{21} R_1 = Q_{11} R_2$. The vacuum
expectation value of $|w_1|^2$ is then $ R_1/Q_{11}=R_2/Q_{21}$
and, by collecting the terms quadratic in $|w_2|$,   the mass 
of  $w_2$  is  found to be zero.
(A similar calculation shows that if we fix the parameters so that 
$w_1$ vanishes in the vacuum  then $w_1$ becomes massless.)
From the point of view of superconducting vortex physics
this result is disappointing: one requires an 
 exponentially  localised $w_2$ condensate 
in order to interpret the unbroken gauge group $U(1)_b$
as the gauge group of electromagnetism. 
If  the condensate
only decays  according to some power law the $U(1)_b$ gauge
invariance is never properly restored outside the core of the vortex
and the  electromagnetic interpretation  is not appropriate.

By adding a suitable perturbation to  the  Lagrangian \suplag\
one could ensure an exponentially localised condensate, of  course
at the expense of destroying the Bogomol'nyi property. 
However, properties such as interactive dynamics and thermodynamics
of such `almost Bogomol'nyi' models can also
be studied with relative ease by  perturbation methods \Shah.
This may be interesting from the point
 of view of the phenomenology of superconducting vortices but  we will
not pursue  it here. Instead we want to exhibit some 
detailed properties of the model \suplag\  by picking a particular 
set of  generators $t_a$.

Thus we return to  the generators \generators, i.e. we set  
 $Q_{11}= Q_{12}=1$
 and $Q_{21}=-Q_{22}=1$  so that 
$r_1=(R_1+R_2)/2$, $r_2=(R_1-R_2)/2$. We also continue with 
the   index saving  terminology   and 
write $a_i = A_i + B_i$ and $b_i = A_i - B_i$, $a=(a_1-ia_2)/2$ and 
$b=(b_1-ib_2)/2$ as well as 
$f_{12} =F_{12} + G_{12}$ and $g_{12} = F_{12} - G_{12}$. Then  
  the Bogomol'nyi 
equations \newhol\ read 
\eqna\twohol
$$
\eqalignno{
(\partial_{\bar z} + i \bar{a})w_1&= 0 & \twohol a  \cr
(\partial_{\bar z} + i \bar{b})w_2&= 0 & \twohol b  \cr
f_{12}  +{1\over 2} (e_1^2 +e_2^2)(r_1- |w_1|^2) + &{1\over 2} (e_1^2-e_2^2)
(r_2 - |w_2|^2) = 0&\twohol c\cr
g_{12}  +{1\over 2}(e_1^2 -e_2^2)(r_1- |w_1|^2) +& {1\over 2}(e_1^2+e_2^2)
(r_2 - |w_2|^2) = 0, &\twohol d\cr
}
$$
and the  generalised vortex equations \elliptic\ are 
\eqna\bivort
$$
\eqalignno{
\Delta \ln |w_1|^2 +(e_1^2 +e_2^2)(r_1- |w_1|^2) + (e_1^2-e_2^2)
(r_2 - |w_2|^2)
 &=4\pi \sum_{s=1}^{n_1} \delta(z-z_{1,s}) & \bivort a
 \cr
\Delta \ln |w_2|^2 +(e_1^2 - e_2^2)(r_1 - |w_1|^2) + 
(e_1^2+ e_2^2)(r_2 - |w_2|^2) &
=4\pi \sum_{t=1}^{n_2} \delta(z-z_{2,t}). & \bivort b \cr
}
$$
We are  interested in  the way the properties of 
solutions, such as energies and magnetic fluxes, depend  on the coupling
constants $e_1$ and $e_2$ and on the values of the parameters
$r_1$ and $r_2$. The latter determine the symmetry pattern.
If $r_1 > 0$ and $r_2 >0 $ the gauge symmetry
is completely broken. Then the fluxes $\varphi_1$ and 
$\varphi_2$  equal the integers $n_1$ and $n_2$  respectively.
Although we do not prove this rigorously here we expect that 
given these integers  there is a   $(n_1+n_2)$ complex parameter
family of solutions of  \bivort{a}\ and \bivort{b}. Each solution
is fully characterised
by the unordered set of  parameters $\{z_{1,1}, ..., z_{1,n_1}\}$
 and $\{z_{2,1}, ..., z_{2,n_2}\}$
which   label the zeros of the scalar fields $w_1$  
and $w_2$. Note that, contrary to the situation for a 
single vortex, the magnitudes of the  magnetic fields  are 
  not necessarily
maximal at these  zeros.

If $r_1 >0$  and  $r_2 =0$, $U(1)_a$
is broken but $U(1)_b$ remains unbroken (and vice-versa for 
$r_1=0$ and $r_2>0$). Thus $\varphi_1$ equals again the integer
 $n_1$ which counts
the zeros of $w_1$, but $\varphi_2$ may have a non-integral part. 
The interpretation, explained in sect. 3, is that $w_2$ has 
a zero at infinity whose order is at least one  but may be non-integral.
Thus only $[\varphi_2 - 1 ]$ zeros (where $[\varphi_2]$ 
denotes the integral part of $\varphi_2$) can be placed arbitrarily
in $\bR^2\cup \{\infty\}$. For given $\varphi_1 =n_1$ and 
$\varphi_2$ we therefore expect there to be 
 a $n_1 + [\varphi_2] -1$ complex
 parameter family of solutions of \bivort{a}\ and \bivort{b}. 

Finally in the special case $r_1=r_2=0$
the full gauge symmetry survives, but we shall see that 
the model has no  finite energy solutions in this case.
 Before we discuss the spectrum of solitons in more detail
we will now show that our model contains, in the 
limit $e_1 \rightarrow \infty$,  the  gauged
$O(3)$ sigma model  introduced in  ref. \Schroers\ and analysed 
further in   ref.  \Yang.

\subsec{The gauged $O(3)$ sigma model revisited}

In  taking   the limit $e_1 \rightarrow \infty$ we 
can now follow the general recipe of  sect. 4. Thus
we impose   simultaneously
the constraint $|w_1|^2 + |w_2|^2 = R_1 >0$. 
Then we solve for $A_{\mu}$ according to \elim:
\eqn\elima{\eqalign{
 R_1 A_{\mu} = iw^{\d}\partial_{\mu} w - w^{\d}t_2 w B_{\mu}.
}}
Since the field $(w_1,w_2)$ is now only defined up to an overall
phase we  can again discuss the model  in terms of the 
ratio $u=w_2/w_1$ \uuu.
In particular the 
 energy functional
for  static fields takes the form
\eqn\sige{\eqalign{
E= {1\over 2} \int d^2 x \,(r_1+r_2) {|D_1 u|^2
+ |D_2 u|^2 \over (1+ |u|^2)^2}  + 
 {1\over e_2^2}G_{12}^2 &+
 e_2^2\left({ r_1|u|^2 -r_2 \over  1 + |u|^2}\right)^2,
}} 
where the covariant derivative on $u$ is $D_i u =
(\partial_i -2 iB_i)u$.
This  functional is a generalised form of the 
energy functional of the gauged $O(3)$ sigma model  introduced
in ref. \Schroers\ 
 and formulated there in terms of  the field  $\bphi$ \pphi.
To recover the formulae in ref.  \Schroers\ one  should set $r_1=1$,  
$r_2=0$ and   $e_2 = 1$,   and 
  identify $-2B_i$  and $-2G_{12}$
with what is called $A_i$ and $F_{12}$  there.
Then the energy functional \sige\ is 4 times the energy functional 
studied in ref.  \Schroers\ and  the magnetic flux 
\eqn\bfl{\eqalign{
\Phi_2 = -{1\over 2\pi} \int d^2x \, G_{12}
}} 
 is ${1\over 4\pi}\times$
the magnetic flux defined in ref.  \Schroers, which is denoted $\Phi$
there.  

Again  we have   the inequality
\eqn\sigb{\eqalign{
E\geq \pi(r_1\varphi_1 + r_2\varphi_2),
}}
and from eq.  \elima\  one finds  the following formulae for the 
fluxes
\eqn\phua{\eqalign{
\varphi_1= {1\over 2\pi i} \int d^2x\, { \overline { D_1 u}D_2 u
-\overline{D_2  u} D_1 u \over (1+|u|^2)^2}
-{1\over 2 \pi} \int d^2 x\, { 2 G_{12} |u|^2 \over 1+ |u|^2 },
}}
and 
\eqn\phub{\eqalign{
\varphi_2= {1\over 2\pi i} \int d^2x\, { \overline{ D_1 u} D_2 u
-\overline{ D_2 u} D_1 u \over (1+|u|^2)^2}
+ {1\over 2 \pi} \int d^2 x\, { 2 G_{12}  \over 1+ |u|^2 }.
}}
Such  integral formulae are useful for explicit  computations,
but to   understand the  geometrical meaning
of  the magnetic fluxes $\varphi_1$ and 
$\varphi_2$ in this model it is best to  recall
how  the magnetic fluxes are related  to the numbers of zeros
of   the scalar fields in the vortex model.
 Since $r_1$ and $r_2$ should  not  both be zero
let us for definiteness assume that $r_1 >0$. Then  $\varphi_1$  is an 
integer (called $n_1$ above) which counts with multiplicity    
the number of zeros of $w_1$ or equivalently the number of poles
of $u$.
The flux $\varphi_2$ counts the zeros of $w_2$  and hence of $u$.
 If  $r_2>0$, i.e.  if  the only remaining 
 $U(1)$ gauge symmetry is spontaneously 
broken,    then  $|u|^2 = r_2/r_1$ at spatial infinity
 and  all zeros of $u$
are at finite $\rho$. Then 
 $\varphi_2$ is also an integer
(called $n_2$ above) which counts these zeros  with multiplicity.
 If $r_2 =0$, i.e. in the case of unbroken
gauge symmetry,  the condition 
 $\varphi_2  \in \bR^{\geq 1}$ tells us that $u$ 
must have a zero at infinity of
order at least 1.

As an aside we note 
 that integer $\varphi_1$ is  called degree of $u$ in  ref. \Schroers.
When $\varphi_2$ is not an integer this is somewhat misleading because 
the scalar field $u$  then decays at spatial infinity according to some
power law with a non-integral exponent. Such a field cannot be 
viewed as a continuous map from  
$\bR^2\cup\{\infty\}$ to  $\bCP^1$ and 
therefore does not have a well-defined degree.
As explained here,  $\varphi_1$ should be thought of more generally 
as the 
number of poles of $u$, counted with multiplicity.  

The Bogomol'nyi equations in this model
may  be derived  either 
by determining the condition for the equality in  \sigb\
to hold, or by subtracting \twohol{a}\ from \twohol{b}\ and 
\twohol{c}\ from \twohol{d}.
In either case the result is
\eqna\Bogu
$$\eqalignno{
(\partial_{\bar{z}}  -2i \bar{B})u &= 0 &\Bogu a\cr
G_{12} +   e_2^2\left({ r_1|u|^2 -r_2 \over  1 + |u|^2}\right)
 &= 0.  &\Bogu b \cr
}
$$
These  equations imply a second  
order elliptic equation   for $u$ which can again be derived 
in two ways. Either, as first shown in ref.
 \Yang, from \Bogu{a}\  and \Bogu{b}\
 via the $\bar \partial$-Poincar\'e lemma 
or  by   subtracting  eq. \bivort{a}\ from eq. \bivort{b}.
In either case  the result is 
\eqn\ellu{\eqalign{
\Delta \ln |u|^2 - 4 e_2^2\left (r_1|u|^2 - r_2 \over 1 + |u|^2\right)
=-4\pi \sum_{s=1}^{n_1}\delta(z-z_{1,s}) +  4\pi \sum_{t=1}^{n_2}
\delta(z-z_{2,t}).
}}
This equation with $r_2=0$ is analysed carefully in ref. \Yang.
It is shown there that in that situation and  for given $\varphi_1$ and 
$\varphi_2$ it has a $\varphi_1 + [\varphi_2]-1$ complex parameter
family of solutions, which agrees with our general
counting  argument given in the discussion of \bivort{a}\ and
\bivort{b}\  above.

 Physically one may  think of  a solution
of the Bogomol'nyi equations \Bogu{a}\  and \Bogu{b}\ 
as a texture carrying 
 magnetic
flux $ 2 \Phi_2 =  \varphi_1 -\varphi_2 $
 which    counts the difference between the number
(with multiplicity)  of poles and the number 
of zeros of $u$.  In the case of spontaneously 
 broken gauge symmetry  the flux is quantised and can be positive,
 negative or zero. In the unbroken case, however, it 
is clear 
from \bfl\ and  \Bogu {b}\  with $r_2 =0$ 
that the magnetic flux $\Phi_2$ is  positive for non-trivial solutions.
Thus we deduce that $\Phi_2$ takes values in a finite interval
\eqn\ineq{\eqalign{
0 <2 \Phi_2 = \varphi_1 -\varphi_2 \leq  \varphi_1 -1.
}}
This formula  was  derived independently by Samols
(as referred to in ref.  \Schroers)  and by Yang in ref. \Yang.
In the next section we shall see that it is  a 
consequence of the general inequality \powercon.

At the end of the introduction 
I mentioned that the wish  better to understand the 
mathematical structure and physical significance
of the gauged $O(3)$ sigma model was  the starting point of 
this paper. In this section we have seen that 
mathematically this model  is 
a limiting case of a certain  gauged linear sigma
  model with Fayet-Iliopoulos
potential terms.  Since the relevant gauged linear sigma model
has solutions which describe  superconducting vortices
of  Bogomol'nyi type  this observation also sheds light on
the physics. When $r_2 =0$  the gauged $O(3)$  sigma model 
has an unbroken $U(1)$ gauge group which we may identify with
the gauge group of electromagnetism. In the core
of a soliton solution the gauge symmetry is broken and in
this sense we may think of the solitons in the gauged $O(3)$ sigma
model  as  superconducting textures. As in 
our  discussion of superconducting vortices  in the previous
section the    Bogomol'nyi property of the model leads to a 
power-law localisation of the 
 scalar condensate, which is phenomenologically disastrous .
However, this problem can again  be solved by moving away from the 
Bogomol'nyi limit. In fact a model of the required type is  studied
in \MaxSkyr. The soliton solutions, called Skyrme-Maxwell solitons
there, are exponentially localised. Asymptotically there is an
unbroken $U(1)$ gauge group which is  broken in the centre of 
a  soliton.
Such a topological soliton may thus properly 
be called  a superconducting
texture. 
These textures only involve one scalar field and one gauge field;
from a mathematical point of view  they therefore  appear to provide 
a more economical model for superconducting topological defects
than   superconducting vortices.

\subsec{Symmetry breaking patterns  and the  soliton spectrum}

The purpose of this subsection is  to highlight 
some consequences of  the 
simple energy formula
\eqn\enbog{\eqalign{
E_{\rm {\footnotefont BPS}}= \pi(r_1 \varphi_1 + r_2 \varphi_2)
}}
for  Bogomol'nyi solitons in the $m=2$ GLSM.
 As we saw above, this formula is valid
for all values  of $e_1$ and $e_2$, including the non-linear 
sigma model limit $e_1 \rightarrow \infty$ (although the interpretation
of $\varphi_1$ and $\varphi_2$  changes  in this limit).
Again we organise the discussion according to 
the symmetry breaking pattern.

First consider the case of completely broken symmetry, i.e. 
$r_1 > 0$ and $r_2 >0 $.
Then fluxes  $\varphi_1$  and $\varphi_2$  take
  arbitrary non-negative integer values  $n_1$ 
and $n_2$.
The resulting spectrum depends in an interesting way on 
whether $r_1/r_2$  is irrational or rational.  In the former case
there is no degeneracy  in the energy level for 
given $\varphi_1$ and $\varphi_2$  other than that coming
from the arbitrary positions of the  $n_1$ and $n_2$ zeros of $w_1$ 
and $w_2$ discussed earlier.
When  $r_1/r_2$ is rational, however, solutions with different 
flux quantum numbers may be  degenerate in energy. More precisely,
writing $r_1/r_2= p/q$, where $p$ and $q$ are two integers
which are relatively prime, we deduce from formula \enbog\ that 
 the solution with $(\varphi_1,\varphi_2) =(n_1,n_2)$ has the same energy
as the solution with $(\varphi_1,\varphi_2) =
(n_1+q,n_2-p)$ (provided these integers are
still positive). In the case $e_1=e_2$, where the 
equations \bivort{a}\ and \bivort{b}\ decouple this degeneracy is
easily understood: if the masses of  a single  $w_1$-vortex and a 
single $w_2$-vortex have a rational ratio, then we expect 
  degeneracies   in the energy levels of superpositions of $w_1$-
and $w_2$ vortices. However, for $e_1\neq e_2$, and particularly
in the limit $e_1\rightarrow \infty$,
this degeneracy  is more  remarkable. In this limit 
we interpret $\varphi_1$ as the degree of $u$ and $2 \Phi_2$
as the magnetic flux carried by the texture;
 in terms of these \enbog\ reads 
\eqn\enbogsig{\eqalign{
E_{\rm {\footnotefont BPS}}= \pi\left(
(r_1+r_2) \hbox{degree}[u] -  2 r_2 \Phi_2\right).
}}
Thus the degeneracy for rational $r_1/r_2$ 
means physically that in the gauged $O(3)$ sigma model
changing the degree of a configuration is   energetically 
 equivalent to changing, by a suitable number of units, 
the magnetic flux it carries.

If $r_1 > 0 $ and $r_2 = 0$ then  
$U(1)_a$ is broken and $U(1)_b$ remains unbroken; in that 
case $\varphi_1$ is still quantised as a non-negative integer,
but $\varphi_2$ may now have a non-integer part.
The energy \enbog\  depends  only on $\varphi_1$ and is degenerate
with respect to changes in $\varphi_2$. Note, however, that  the range
of  $\varphi_2$  is restricted  by  the  finite-energy condition
$\varphi_2 \geq 1$ and  the  further constraint coming  from \powercon\
for non-vanishing solutions.
In this case it  reads
\eqn\interest{\eqalign{ 
\varphi_2 < {e_1^2 - e_2^2 \over  e_1^2 +e_2^2} \varphi_1.
}}
Together these conditions   force   $\varphi_2$ 
to lie in  a finite interval. In particular
they may force $\varphi_2$  to vanish 
 for certain values of $e_1$ and $e_2$.
This holds in  particular for $e_1=e_2$ where this result, too, 
is easily understood. The  equation \bivort{b}\ 
for $w_2$ is then  just the abelian Higgs vortex equation \abhivort.
As mentioned in sect. 3   this equation has no solution
other than  the 
trivial solution when $r=0$. In the non-linear 
sigma model  limit $e_1 \rightarrow \infty$
the formula \interest\  yields the promised re-derivation
of \ineq. It now reads $\varphi_2 < \varphi_1$, which we combine with
the earlier condition $\varphi_2 \in \bR^{\geq 1}$ to 
\eqn\intimp{\eqalign{
1\leq \varphi_2  < \varphi_1.
}}
This  is   equivalent  to the inequality  \ineq. Note in particular that 
this condition cannot be satisfied if $\varphi_1=1$ and that 
there is therefore no solution  with precisely one pole  in the gauged 
$O(3)$ sigma model. This result was first derived for 
the spherically symmetric case in  ref. \Schroers\ and proved 
in general in ref. \Yang.

The reverse situation, with $r_1 =0$ and $r_2 > 0$ can
be discussed in an  analogous fashion, and the case 
$r_1=r_2=0$ does not lead to any interesting solution:
now  the condition \powercon\  implies, for non-trivial solutions,
 \interest\ as well as 
\eqn\aaa{\eqalign{
\varphi_1 < {e_1^2 - e_2^2 \over  e_1^2 +e_2^2} \varphi_2
}}
which is incompatible with \interest.
Thus the trivial solutions $w_1=w_2=0$ is the only finite 
energy solution
of the equations \bivort{a}\ and \bivort{b}\ in this case.

\newsec{Abelian Chern-Simons models of Bogomol'nyi type}

Consider the following   Lagrangian density  for a theory where
the dynamics of all the gauge fields are governed by  Chern-Simons
terms:
\eqn\aaa{\eqalign{
{\cal L}_{CS} = -{1\over 2}
 (D_{\mu}w)^{\d}D^{\mu}w -
\sum_{a=1}^m {1\over 2\kappa_a} A^a_{\mu}
 \partial_{\nu}A^a_{\lambda}\epsilon^{\mu\nu\lambda} -
V_{CS}(\kappa_a, w, R_a),
}}
with $m$  dimensionless  coupling constants $\kappa_a$.
 In this section 
we want to determine the  potentials $V_{CS}$
 such  that the model has static soliton solutions of 
Bogomol'nyi type.

The Chern-Simons term is independent of the metric,  so it
does not contribute to the energy-momentum tensor. As a 
result the energy functional reads
\eqn\aaa{\eqalign{
E_{CS}= {1\over 2} \int d^2x (D_0 w)^{\d}D_0w + (D_1 w)^{\d}D_1w +
 (D_2 w)^{\d}D_2w + V_{CS}.
}}
However, the energy does depend on the gauge fields through the covariant
derivatives.
In particular the time-component of the gauge fields is
 determined as a function of the scalar fields and the spatial
components  of the field strength by  the Chern-Simons
version of Gauss's law:
\eqn\aaa{\eqalign{
j^a_0 ={1\over \kappa_a} F^a_{12},
}}
where $j^a_{\mu}$ is defined as in \current.
This constraint means that, even for static fields we cannot
set $A^a_0 =0$. Instead we find, for  time-independent $w_a$,
\eqn\aaa{\eqalign{
\sum_{b=1}^m w^{\d}t_a t_b w \,  A^b_0 = {1\over \kappa_a} F^a_{12}.
}}
Thus the energy functional $E_{CS}$ can be expressed entirely in
terms of spatial components of the gauge fields and the  spatial
part of the field strengths. To do this it is convenient to 
introduce  the notation $\bF$ for the column vector with components
$F^a_{12}/\kappa_a$, $\bV$ for the column vector with
components $\kappa_a(R_a - w^{\d}t_aw)$ and $\bW$ for the $m\times m$
matrix with  components $W_{ab} = w^{\d}t_at_bw$, 
 $a,b =1,..,m$. We claim that 
by setting 
\eqn\aaa{\eqalign{
V_{CS} = {1\over 8}  \bV^t \bW \bV = {1\over 8}  \sum_{a,b =1}^m
 \kappa_a(R_a - w^{\d}t_a w )\,  w^{\d}t_at_b w \,\,
 \kappa_b(R_b - w^{\d}t_bw)
}}
we obtain an energy functional of Bogomol'nyi type:
\eqn\aaa{\eqalign{
E_{CS} &= {1\over 2} \int d^2 x (D_1w)^{\d}D_1w + (D_2w)^{\d}D_2w  + 
\bF^t \bW^{-1}\bF
+{1\over 4}  \bV^t \bW \bV  \cr
&= {1\over 2}\int d^2 x |(D_1 \pm i D_2)w|^2 + (\bF \pm {1\over 2} 
 \bW \bV)^t 
\bW^{-1} (\bF \pm {1\over 2} \bW \bV) \cr
& \mp {1\over 2}  \sum_{a=1}^m R^a \int d^2x F^a_{12}.
}}
Here the superscript $t$ means transposition and 
 we have again  used the boundary condition \boundb\
to integrate by parts and exploited the identity \curvature.
As  in sect.  2  we therefore deduce the inequality 
\eqn\aaa{\eqalign{
E_{CS} \geq \pi |T|
}}
with equality if and only if  one of the Bogomol'nyi equations holds
\eqna\cshol
$$\eqalignno{
(D_1 \pm i D_2)w &=0 &\cshol a  \cr
{F_{12}^a   \over  \kappa_a} \pm {1\over 2} 
 \sum_{b=1}^m w^{\d} t_a t_b w\,  \kappa_b 
 (R_b - w^{\d} t_b w) & = 0. &\cshol b \cr
}
$$

In the case $m=1$ the general prescription we have discussed  
leads to the abelian Chern-Simons-Higgs model with  a sextic
potential,  which has been much discussed in the literature.
 One reason why this model has attracted 
attention is  that it contains both topological and non-topological
solitons \CSHiggs. 
The generalised models described here include, 
for $m=2$, the choice of generators \generators\ and $\kappa_2=0$, 
the semilocal Chern-Simons model discussed in ref. \semiCS, 
and have in general
  a variety of both topological and non-topological soliton solutions
whose properties deserve to be studied further.
However, here we  now turn to   another point of principal interest.

Like in the case of GLSM's with the gauge field governed by a 
Maxwell term it is interesting to consider limits in which  some 
of the gauge fields are eliminated, thus leading to  a (possibly
gauged)  non-linear sigma model. Specifically we can eliminate
the Chern-Simons terms for  the gauge  fields $A_{\mu}^a$,
where $a$ runs over the subset of indices 
$I=\{1, ..., d\}$, $d < m$, 
by taking the limit
\eqn\CSlimita{\eqalign{
\kappa_a \rightarrow \infty, \qquad a\in I
}}
and simultaneously imposing
\eqn\CSlimitb{\eqalign{
(R_a -w^{\d} t_aw) = 0,  \qquad a\in I.
}}
Then  the  equations of motion for the gauge fields $A_{\mu}^a$,
$a\in I$,  are  again \elimaa, allowing us to express these
gauge fields  in terms of the scalar fields and the complementary
set of gauge fields:
\eqn\cselim{\eqalign{
\sum_{b=1}^d w^{\d} t_a t_b w \, A^b_{\mu} = 
i w^{\d} t_a ( \partial_{\mu} + i \sum_{b=d+1}^m t_b A_{\mu}^b) w.
}}

However, there is a subtlety here:  in the Chern-Simons case,
unlike in the Maxwell case, simply taking the value 
of the potential  in the limit \CSlimita\ and \CSlimitb\ as the  
potential for the remaining degrees of freedom does not appear  
to lead to  a model with interesting   Bogomol'nyi solitons.
The quickest way  to see this is  naively to apply the limit
\CSlimita\ and \CSlimitb\  to the Bogomol'nyi equation \cshol{b}.
The equations labelled by $a\in I$ then lead to purely algebraic
constraints on the  scalar fields which eliminate many 
solutions. Here we therefore follow a different path 
and  define a new potential as follows. 
Let $\tilde I =\{d+1, ... ,m\}$ be the complementary index set
to $I$ and let $\tilde \bF$ and $\tilde \bV$ be the column vectors
with components $F^{\tilde a}_{12}/\kappa_{\tilde a}$ and 
$\kappa_{\tilde a}(R_{\tilde a} - w^{\d}t_{\tilde a}w)$ respectively,
where $\tilde a \in \tilde I$. Further  write $\widetilde{\bW^{-1}}$
for the matrix  with elements $(\bW^{-1})_{\tilde a\tilde b}$,
$\tilde a, \tilde b \in \tilde I$.
Then the kinetic term for the gauge field in the Chern-Simons energy
functional in the limit \CSlimita\ is 
\eqn\aaa{\eqalign{
 \tilde \bF^t \widetilde{\bW^{-1}} \tilde \bF
}}
and thus the  appropriate
potential is
\eqn\CSpot{\eqalign{
V_{\widetilde{CS} }= {1\over 8} 
\tilde \bV^t \left( \widetilde{\bW^{-1}}\right)^{-1}
\tilde \bV.
}}
With this choice the
 energy functional  
\eqn\aaa{\eqalign{
E_{\widetilde{CS}}= {1\over 2} \int d^2 x\,
 (D_1w)^{\d}D_1w + (D_2w)^{\d}D_2w  + 
\tilde \bF^t \widetilde{ \bW^{-1}} \tilde \bF
+ {1\over 4}  \tilde \bV^t \left( \widetilde{\bW^{-1}}\right)^{-1}
\tilde \bV
}}
 satisfies
\eqn\aaa{\eqalign{
E_{\widetilde{CS}} \geq \pi |T|
}}
with equality if and only if  one of the Bogomol'nyi equations holds
\eqna\csehol
$$\eqalignno{
(D_1 \pm iD_2)w &=0 &\csehol a  \cr
{ F_{12}^{\tilde a}  \over  \kappa_{\tilde a}} \pm {1\over 2}
  \sum_{\tilde {b}=d+1}^m( \widetilde{\bW^{-1}})^{-1}_{\tilde a \tilde b}
 V_{\tilde b}     & = 0,\qquad \tilde{a}\in \tilde{I}. &\csehol b
}
$$
In the first of these equations the gauge potentials $A_i^a$, $a\in I$
are again determined by \cselim.

The potential \CSpot\ is, in general,
of a  more complicated  form  than the  sextic polynomial 
that is familiar in Chern-Simons theories of Bogomol'nyi type.
Thus it may be useful to 
 illustrate the general  discussion by  considering   again the 
case  $m=2$, with  
entire maximal torus of $U(2)$ gauged. We will again use the 
index saving notation of sect. 5.

With $t_1$ and $t_2$ as  in \generators,  the matrix $\bW$
is 
\eqn\aaa{\eqalign{
\bW  = \pmatrix{w^{\d}w & w^{\d}t_2 w  \cr w^{\d}t_2 w  & w^{\d}w }
}}
so that   in the limit $\kappa_1 \rightarrow \infty$
and with $R_1 - w^{\d} w =0$, the matrix $\widetilde{ {\bW}^{-1}}$
is just a number: 
\eqn\aaa{\eqalign{
\widetilde{ {\bW}^{-1}} = {R_1 \over R_1^2 - (w^{\d}t_2 w)^2 }.
}}
Thus the Chern-Simons potential is 
\eqn\pott{\eqalign{
V_{CS} = {\kappa^2_2\over 8 R_1}
(R_1^2 - (w^{\d} t_2 w)^2)(R_2 - w^{\d}t_2 w)^2
}}
and the Bogomol'nyi equations are
\eqna\bogo
$$\eqalignno{
(D_1 \pm iD_2)w  &= 0 & \bogo a \cr
 G_{12} \pm   {\kappa_2^2\over 2 R_1}
(R_1^2 - (w^{\d} t_2 w)^2)(R_2 - w^{\d}t_2 w)
& = 0. &\bogo b
}
$$
Formulating this model in terms of the field $u$ \uuu\
is again instructive. Using also again 
 the parameters $r_1$ and $r_2$  defined in  
sect.  5,  the potential  \pott\   takes the form
\eqn\pottt{\eqalign{
V_{CS} = 2\kappa_2^2(r_1+r_2){ |u|^2 (r_1|u|^2 -r_2)^2 \over (1+|u|^2)^4}.
}}
Choosing the upper sign in the Bogomol'nyi equations they 
now read  
\eqna\bogok
$$\eqalignno{
(\partial_{\bar z} - 2i\bar B)u &=0&\bogok a \cr
G_{12} + 4 \kappa_2^2 (r_1+r_2)
 {|u|^2( r_1|u|^2 -r_2) \over (1+|u|^2)^3 } &=0. &\bogok b \cr
}
$$
In fact these equations and their soliton  solutions are  studied
with slightly different notation   in  ref. \Gosh\  and  ref. \Tigran\
for the case of unbroken $U(1)$ gauge symmetry $r_2 =0$. 
In  ref. \Lee\  the case of broken gauge symmetry, $r_2 > 0$ 
is also considered. 
As often in Chern-Simons theories there are both topological
and non-topological solitons in this model. All soliton solutions
carry magnetic flux, and in the case of broken gauge symmetry
the topological solitons may derive their topological stability
from the (quantised) magnetic flux, from the number of poles of $u$
(the ``degree'') or from a combination of both. For more details
we refer the reader to the papers \Gosh, \Tigran\ and \Lee\ and 
also to ref.  \Jens, where a related model is discussed. 
However,  it  should be clear from this brief case study that the 
family of gauged linear sigma models with Chern-Simons  terms defined 
in this section  have an even richer spectrum of solitons than
their Maxwell cousins.

\newsec{Discussion and Outlook}

In (2+1) dimensions, 
many interesting topological solitons of Bogomol'nyi
type can be studied in  a  unified  way in terms of the
gauged linear sigma models  discussed in this paper. 
Thinking of the GLSM's with Maxwell term and Fayet-Iliopoulos
potential terms as a generalisation of the abelian Higgs
model and keeping in mind the phenomena encountered here, 
 it would be interesting to address the following points
in full generality.

Can one extend the 
analysis of  ref. \bogvor\ and ref. \Yang\ to 
 establish  rigorous general existence and uniqueness theorems 
for the   generalised  vortex equation \elliptic?
If   in the 
general model discussed in sect. 3 the gauge symmetry
is completely broken (i.e.  if all $r_{\alpha}>0$) 
 then 
all the fluxes $\varphi_{\alpha}$  are non-negative integers
and the examples studied above  suggest that for given fluxes 
there is a $\sum_{\alpha =1}^m \varphi_{\alpha}$  complex
parameter family of solutions. If the gauge symmetry is only 
partially broken, say $r_{\alpha} > 0$ for $1\leq \alpha \leq d$
and $r_{\alpha} =0$ for $d < \alpha \leq m$, then I conjecture that 
 there 
is generically  a $\left( \sum_{\alpha =1}^d  \varphi_{\alpha} + 
\sum_{\alpha =d+1}^m [\varphi_{\alpha} - 1]\right)$ complex parameter family
of solutions for given fluxes. However, in this case the 
fluxes also have to  satisfy the constraints \powercon.
Here we have only shown that these constraints are necessary 
conditions for finite-energy solutions. It would be interesting
to know whether they (or possibly  some stricter version of them) are also
sufficient for the existence of finite energy solutions.

Having counted the solutions it is then interesting  to look at
the structure - differentiable and metric -  of the  
moduli spaces of solutions. The examples studied here 
suggest that the moduli spaces of vortices on $\bR^2$ 
are typically diffeomorphic to $\bC^D$ for some $D$ (determined 
by the above counting arguments) and have smooth metrics.
However, in the  limit $e_a \rightarrow \infty$ for 
some $a\in\{1,...,m\}$,
where vortices turn into (possibly gauged) textures a number of 
interesting things happen. The interpretation of the  moduli
changes. Whereas vortex moduli simply characterise the (unordered)
zeros of the scalar fields, the moduli for textures include
internal  as well as  position parameters.
Taking certain limits of the internal parameters corresponds
to making the texture infinitely spiky, and these limiting 
values are  therefore  forbidden as moduli  for textures. Thus
the moduli spaces for textures  are typically obtained from 
those of vortices by removing certain  
lower dimensional algebraic submanifolds.
Metrically, the moduli spaces of textures are also   worse
behaved than those of vortices. The example of the $\bCP^1$
lumps suggest that non-complete and non-finite  metrics
arise. In studying moduli spaces of textures it is 
therefore  useful to bear in mind that they
may be thought of  as singular limits of smooth vortex moduli
spaces. In that way, singularities  may be understood and,
if desired, circumvented.

Some of the  questions raised so far  are   completely answered in 
the case where  GLSM's  are defined  on compact Riemann surfaces.
There the  counting of solutions  can be done using  standard theorems 
in algebraic geometry, and much can be deduced about the 
 topological and (complex)
differentiable structure of the  moduli spaces from the key  
observation that they (like the target spaces of non-linear 
sigma models encountered in sect. 4, and for similar reasons)
are  toric varieties; see ref. \MP\ for a pedagogical explanation of this
statement 
 and for  further references. In particular the intersection 
ring of the moduli spaces is described  in the literature, see
again \MP. 
This is of interest from the point of view of vortex physics
because knowing the  intersection   ring
of the moduli spaces  is the crucial ingredient in  the 
study of the statistical mechanics of vortices in ref.   \statmech.
There  the statistical mechanics of Nielsen-Olesen  vortices 
is studied  on (amongst other Riemann surfaces) the torus, 
which is equivalent to studying vortices on  the plane with
periodic boundary conditions. Thus it appears that  much of 
the information needed similarly to study  the statistical
mechanics of the more 
general types of vortices described here  can be found 
in the mathematical and string theory literature. However,  one 
important ingredient in the analysis of ref. \statmech,
  a certain normalisation
factor,  is lacking. Computing it requires  some knowledge of 
the Riemannian structure that the moduli space inherits from the
kinetic energy of the field theory. Since  that  metric 
is only natural form a (2+1)-dimensional point of view it
has not been considered  in the string theory context. 

It  is also interesting to ask some of the above 
questions  for the
 Chern-Simons version of the GLSM's  introduced in
this paper.  Already the first step, the general  analysis of the
soliton spectrum,   is more complicated in this case. 
There are now both topological and non-topological solitons
and since  the scalar  potential no longer has a universal
polynomial structure the finite energy condition needs to be 
analysed for  each model separately.  It is also
natural to contemplate   further generalisations, such
as   mixed models, where
some of the $U(1)$ gauge fields are governed by Maxwell terms
and others by Chern-Simons terms.  It should even be possible
to  construct general  abelian  GLSM's  of Bogomol'nyi type where  
both a Maxwell and a Chern-Simons term is included for each gauge
field. As shown first  for the abelian Higgs model  
in ref. \LLM,  later extended in ref. \Kim\ and  recently 
demonstrated  in the context of the gauged $O(3)$ sigma model
in refs. \Gosh\ and \Lee,  this requires the introduction of additional
neutral scalar fields. In such models the gauge fields can
get their masses from the Chern-Simons term (``topological mass'')
or through the Higgs mechanism.
The analysis in ref. \Kim\  suggests that generalising this 
construction
for  abelian GLSM's   will lead to  models which 
can   accommodate a truly bewildering
array of massless, massive, topological and non-topological 
particles. Thus it seems that the framework of abelian  GLSM's  
 provides an extremely versatile  kit for the 
construction of Bogomol'nyi type field theories in (2+1)
dimensions. The Maxwell version mainly studied here also
shows that this kit  comes  equipped with powerful tools for 
exploring the physics of these field theories.

\ack{
I   thank Robbert Dijkgraaf and   Jae-Suk Park for  discussions
about the   connection between  GLSM's and toric varieties  
and acknowledge financial support through a  Pioneer Fund of 
the Nederlandse Organisatie voor Wetenschappelijk Onderzoek (NWO). 
}

\listrefs
\bye